\documentclass[prd,reprint,nofootinbib,showpacs,superscriptaddress]{revtex4-1}
\usepackage{graphicx} 
\usepackage{hyperref}
\usepackage{amsfonts}
\usepackage{amsmath,amssymb}
\usepackage{bm} 
\usepackage{color}
\usepackage{epstopdf} 
\usepackage{epsfig}
\usepackage{subfig} 
\usepackage{float}

{\rm }

\def\be{\begin{equation}}
 \def\ee{\end{equation}}
 \def\bea{\begin{eqnarray}}
 \def\eea{\end{eqnarray}}
 \def\bes{\begin{eqnarray}}
 \def\ees{\end{eqnarray}}
 \def\bi{\begin{itemize}}
 \def\ei{\end{itemize}} 

\def\2{\frac{1}{2}}
\def\4{\frac{1}{4}}

\begin{document}

\title{Experimental passive state preparation for continuous variable quantum communications}

\author{Bing Qi}
\email{qib1@ornl.gov}
\affiliation{Quantum Information Science Group, Computational Sciences and Engineering Division, Oak Ridge National Laboratory, Oak Ridge, TN 37831, USA}
\affiliation{Department of Physics and Astronomy, The University of Tennessee, Knoxville, TN 37996, USA}

\author{Hyrum Gunther}
\affiliation{Department of Electrical and Computer Engineering, Brigham Young University, Provo, UT 84602, USA}

\author{Philip G. Evans}
\affiliation{Quantum Information Science Group, Computational Sciences and Engineering Division, Oak Ridge National Laboratory, Oak Ridge, TN 37831, USA}

\author{Brian P. Williams}
\affiliation{Quantum Information Science Group, Computational Sciences and Engineering Division, Oak Ridge National Laboratory, Oak Ridge, TN 37831, USA}

\author{Ryan M. Camacho}
\affiliation{Department of Electrical and Computer Engineering, Brigham Young University, Provo, UT 84602, USA}

\author{Nicholas A. Peters}
\affiliation{Quantum Information Science Group, Computational Sciences and Engineering Division, Oak Ridge National Laboratory, Oak Ridge, TN 37831, USA}
\affiliation{Center for Interdisciplinary Research and Graduate Education, The University of Tennessee, Knoxville, Tennessee 37996, USA}

\date{\today}
\pacs{03.67.Dd}

\begin{abstract}

{In the Gaussian-modulated coherent state quantum key distribution (QKD) protocol, the sender first generates Gaussian distributed random numbers and then encodes them on weak laser pulses \emph{actively} by performing amplitude and phase modulations. Recently, an equivalent \emph{passive} QKD scheme was proposed by exploring the intrinsic field fluctuations of a thermal source [B.~Qi, P.~G.~Evans, and W.~P.~Grice, {\em Phys. Rev. A} {\bf 97}, 012317 (2018)]. This passive QKD scheme is especially appealing for chip-scale implementation since no active modulation is required. In this paper, we conduct an experimental study of the passively encoded QKD scheme using an off-the-shelf amplified spontaneous emission source operated in continuous-wave mode. Our results show that the excess noise introduced by the passive state preparation scheme can be effectively suppressed by applying optical attenuation and secure key can be generated over metro-area distances. \footnote{This manuscript has been authored in part by UT-Battelle, LLC, under contract DE-AC05-00OR22725 with the US Department of Energy (DOE). The US government retains and the publisher, by accepting the article for publication, acknowledges that the US government retains a nonexclusive, paid-up, irrevocable, worldwide license to publish or reproduce the published form of this manuscript, or allow others to do so, for US government purposes. DOE will provide public access to these results of federally sponsored research in accordance with the DOE Public Access Plan (http://energy.gov/downloads/doe-public-access-plan).
}}

\end{abstract}

\maketitle

\section{Introduction}
\label{sec:1}

Quantum key distribution (QKD) allows two remote users (Alice and Bob) to establish a secure key by transmitting quantum states through an untrusted channel \cite{Gisin02, Scarani09, Lo14, Diamanti16}. The generated secure key can be further applied in various cryptographic protocols to achieve long-term proven security against adversaries with unlimited computational power.

QKD protocols are commonly divided into two families based on encoding schemes: discrete-variable (DV) QKD or continuous-variable (CV) QKD. In contrast, classical optical communications are typically grouped into two categories based on detection schemes: direct detection or coherent detection. In principle, combining both encoding (DV or CV) and detection (direct or coherent) could lead to four families of QKD protocols. Among them, DV-QKD with direction detection (single photon detection) \cite {BB84, E91} and CV-QKD with coherent detection (optical homodyne detection) \cite{Ralph99, Hillery00, GMCSQKD} are dominant, although other protocols, such as CV-QKD using single photon detection \cite{Qi06}, do exist. For simplicity, in this paper we use the term CV-QKD to refer to CV-QKD using coherent detection.

CV-QKD's most distinguishing feature is coherent detection.  It enables high speed optical homodyne detection with no dead time which could give high secure key rates over short distances. Further, the intrinsic filtering provided by the local oscillator in a coherent receiver can effectively suppress background noise and enable QKD through conventional dense wavelength division multiplexed fiber networks in the presence of strong classical traffic \cite{Qi10, Kumar15, Eriksson19}.  The similarity between a CV-QKD system and a classical coherent communication system opens the door for simultaneous quantum and classical communications \cite{Qi16} with a technological pathway towards fully integrated, on-chip, photonic implementation \cite{Zhang19}. Integrating CV-QKD on a chip may have several benefits.  It allows for the integration of multiple photonic functions into a single compact circuit.  In particular, the phase-sensitive optical circuits commonly used for CV-QKD can be made more robust to temperature-induced phase drifts by reducing path-length differences on chip.  Furthermore, silicon photonic devices are compatible with complementary metal-oxide-semiconductor (CMOS) processes that enable monolithic integration of electronics and photonics, potentially leading to significant cost reduction which would enable the wide-spread utilization of QKD.  

One important CV-QKD protocol is the Gaussian-modulated coherent states (GMCS) QKD protocol \cite{GMCSQKD}, which has been implemented with standard off-the-shelf telecom components, such as laser sources, optical homodyne detectors, and optical intensity and phase modulators \cite{Lodewyck07, Qi07, Jouguet13, Huang16, ZLC19}. In the GMCS protocol, Alice first generates Gaussian distributed random numbers $x_A$ and $p_A$ and then prepares a coherent state $|x_A+ip_A\rangle$ sent to Bob through a channel controlled by the adversary (Eve). The quantum state preparation is commonly implemented \emph{actively} using amplitude and phase modulation. High speed modulation with high extinction is extremely challenging in chip-scale silicon photonics. While on-chip modulation with extinction ratio above 65 dB has been demonstrated recently \cite{Liu2017}, the high speed on-chip modulators required for active QKD encoding adds significant cost, manufacturing time and complexity, and are the principal source of loss in most integrated photonic circuits.  Therefore, the potential to remove the modulators used for encoding may yield significant reductions in cost, manufacturing time, and on-chip loss.

One could simplify the chip-scale implementation by adopting a passive CV-QKD protocol, where the amplitude and phase modulators in the GMCS QKD are replaced by a thermal source, beam splitters, optical attenuators and homodyne detectors \cite{Qi18}. As we have shown in \cite{Qi18}, given that Alice's QKD transmitter is trusted, the passive CV-QKD protocol is equivalent to GMCS QKD. This means that the well-established security proofs for GMCS QKD can be applied to passive CV-QKD directly. More recently, this passive state preparation scheme has also been extended to measurement-device-independent CV-QKD \cite{Bai19}. It could also be applied in other CV quantum communication protocols, such as quantum secret sharing \cite{Grice19} and quantum digital signatures \cite{Croal16}.

In this paper, we conduct experimental studies of passive CV-QKD using a practical multi-mode thermal source. As we will show below, the excess noise due to the passive state preparation scheme can be effectively suppressed by Alice applying optical attenuation and, secure key can be generated over metro-area distances using an off-the-shelf amplified spontaneous emission source operated in continuous-wave (cw) mode. 

This paper is organized as follows: in Sec. II, we review the passive CV-QKD protocol and compare it with conventional GMCS QKD \cite{GMCSQKD, Weedbrook04} as well as entanglement-based CV-QKD using a two-mode squeezed vacuum state \cite{Grosshans03}. In Sec. III, we present our experimental setup and develop a corresponding noise model. In Sec. IV, we present the experimental results. Finally, we conclude this paper with a brief summary in Sec. V.

\section{The protocol and its security}
\label{sec:2}

In the GMCS QKD protocol \cite{GMCSQKD}, Alice generates Gaussian distributed random numbers $x_A$ and $p_A$ using a trusted random number generator, prepares a coherent state $|x_A+ip_A\rangle$ accordingly, and transmits it to Bob, as shown in Fig. 1(a). At Bob's end, he can either measure a randomly chosen quadrature of the incoming quantum state by conducting single homodyne detection \cite{GMCSQKD} or measure both quadratures simultaneously by conducting conjugate homodyne detection (which is called the heterodyne protocol) \cite{Weedbrook04}. Alice and Bob further estimate channel-induced noise and other QKD parameters by comparing a subset of their data through an authenticated classical channel. This allows them to upper bound the information that could be gained by a third party Eve. Given the correlation between Alice and Bob's data is above certain threshold, they can perform reconciliation and privacy amplification to generate a final secure key.

There are different reconciliation algorithms in GMCS QKD. In this paper, we consider the heterodyne protocol \cite{Weedbrook04} using reverse reconciliation \cite{GMCSQKD} where Alice tries to determine Bob's measurement results from her data. In this case, secure key can be generated as long as the mutual information between Alice and Bob is larger than the information Eve could have on Bob's measurement results. More details about the GMCS QKD can be found in recent reviews \cite{Diamanti15, Laudenbach18}.

\begin{figure}[t]
	\includegraphics[width=.45\textwidth]{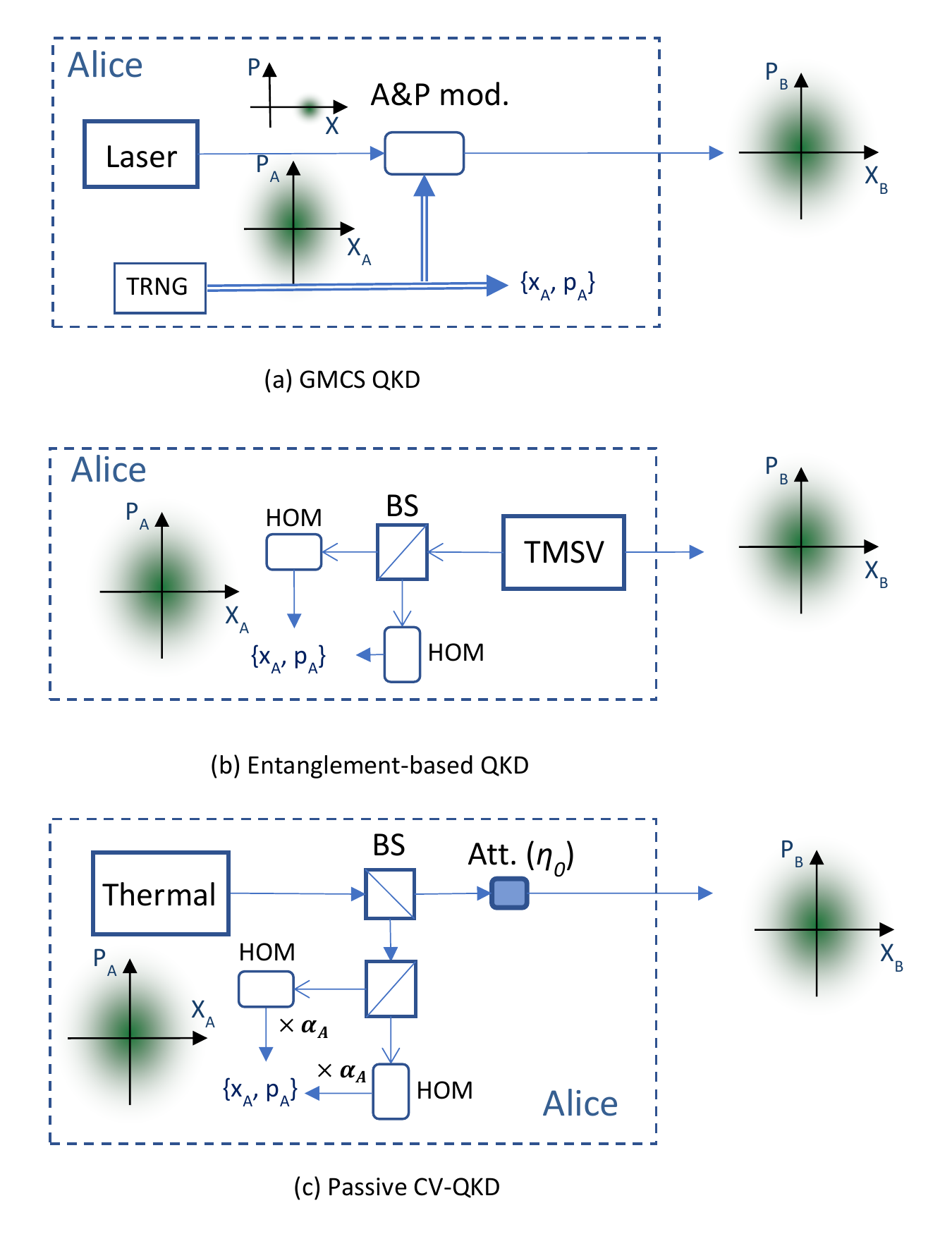}
	\captionsetup{justification=raggedright,
					singlelinecheck=false }
	\caption{Three equivalent quantum state preparation schemes. (a) The Gaussian-modulated coherent states (GMCS) QKD \cite{GMCSQKD}. TRNG, true random number generator; A$\&$P mod., amplitude and phase modulators. (b) Entanglement-based CV-QKD \cite{Grosshans03}. TMSV, two-mode squeezed vacuum state; HOM, homodyne detector; BS, beam splitter. (c) Passive CV-QKD \cite{Qi18}. Att., optical attenuator.} 
	\label{fig:1}
\end{figure}

Note from Eve's point of view, the quantum state from Alice is a mixture of all possible coherent states, which is simply a thermal state with an average photon number $n_0=V_A/2$, where $V_A$ is Alice's modulation variance. We remark that throughout this paper, all the noise variances and modulation variances are defined in  shot-noise units. There are different ways (corresponding to different QKD protocols) for Alice to prepare the outgoing thermal state, as shown in Fig. 1. As long as Alice's QKD transmitter is within a trusted location, Eve cannot tell which protocol is actually carried out by Alice. This suggests all these QKD protocols are equivalent in terms of security. In fact, the security of the GMCS QKD is commonly analyzed based on an entanglement-based CV-QKD protocol using a two-mode squeezed vacuum state \cite{Grosshans03}, as shown in Fig. 1(b). In this entanglement-based protocol, by performing conjugate homodyne detection on one mode of a properly chosen two-mode squeezed vacuum state, Alice acquires Gaussian distributed random numbers $x_A$ and $p_A$ with the desired variance of $V_A$. In the meantime, Alice transmits the other mode, which is projected to a coherent state $|x_A+ip_A\rangle$, in her perspective, to Bob through the quantum channel. From Eve's point of view, since she has no information about $x_A$ and $p_A$, the state from Alice is thermal, as in the case of GMCS QKD. 

In passive CV-QKD \cite{Qi18}, the intrinsic field fluctuations of a thermal source are utilized to generate secure key, as shown in Fig. 1(c). In this protocol, Alice splits the output of a thermal source into two spatial modes using a beam splitter. One mode is sent to Bob after being attenuated using an optical attenuator, while the other mode will be measured locally by Alice using conjugate homodyne detection. Alice further numerically scales down her measurement results by a factor of $\alpha_A$ acquiring Gaussian-distributed random numbers $x_A$ and $p_A$, her estimation of  the quadrature values of the outgoing mode (see Section III for the details about how to determine the optimal $\alpha_A$). By choosing a proper combination of source intensity and optical attenuation, the quadrature variance of the outgoing mode can be set to the desired value $V_A$. Again, from Eve's point of view, she cannot distinguish the quantum state sent in passive CV-QKD from the one in the GMCS QKD (or, for that matter, entanglement-based CV-QKD), so all three protocols are equivalent in terms of security. We remark that the combination of the balanced beam splitter and the optical attenuator in Fig. 1(c) can be replaced by a single unbalanced beam splitter with a suitable splitting ratio. The latter configuration can lead to a more efficient use of the thermal source.

While the security is not dependent on which of the above schemes is employed, the secure key rate is sensitive to the excess noise generated in the quantum state preparation process. Since the quantum states sent by Alice are identical in the three schemes, so is the information Eve could gain on Bob's measurement results. The mutual information between Alice and Bob directly connects to Alice's uncertainties on the quadrature values of the outgoing mode. In GMCS QKD (and the above entanglement-based CV-QKD), the minimum uncertainty Alice could achieve on either quadrature of the outgoing mode is equal to one, as determined by the uncertainty principle in quantum mechanics. In Appendix A, we present a detailed noise model of the passive CV-QKD  protocol and show that Alice's uncertainty on the outgoing mode is given by
\bes\label{eq1} \Delta=\dfrac{2V_A(\upsilon_{a}+1)}{V_A \eta_{a}+2\eta_0(\upsilon_{a}+1)}\eta_0+1, \ees
where $\eta_0$ is the transmittance of the optical attenuator, $\eta_{a}$ and $\upsilon_{a}$ are the efficiency and noise variance of Alice's detector, respectively. See Eq. (A5) in Appendix A.

From Eq. (1), the excess noise due to the passive state preparation scheme (which equals $\Delta-1$) can be effectively suppressed by introducing optical attenuation at Alice (while keeping $V_A$ at the desired value). Equation (1) suggests that regardless of the amount of detector noise, as $\eta_0$ approaches zero, so does the excess noise. In practice, this is convenient, since Alice's detector does not need to be shot-noise limited. To further illustrate the role of the optical attenuation, in Appendix A, we explicitly show that under the beam-splitting attack, the information advantage between Alice and Bob, when compared to that between Eve and Bob, can be improved by introducing optical loss at Alice. 

In practice, $\eta_0$ cannot be zero, otherwise, to have non-zero average photon number in the outgoing mode, the intensity of the thermal source has to be infinite. Security proofs of practical CV-QKD, taking into account the excess noise contributed by both Alice and Bob, are well developed \cite{Diamanti15, Laudenbach18}. In next section, we will present our experimental setup, develop a corresponding noise model, and quantify the QKD performance using the established security proof.

\section{Experimental setup and noise model}
\label{sec:3}

The experimental setup is shown in Fig. 2. A fiber amplifier (PriTel, Inc.) operated in cw mode (Amplified Spontaneous Emission: ASE in Fig. 2) with vacuum state input is employed as a broadband thermal source. Previous studies have shown that the output from an unseeded fiber amplifier is thermal \cite{WH98, VV00, Qi18, Qi17}. To select a single polarization mode, a fiber pigtailed polarizer (Pol in Fig. 2) is placed right after the light source. Since the spectral bandwidth of the source is about 30 nm, a 0.8 nm optical bandpass filter centered at 1542 nm (BP in Fig. 2) is employed to reduce the optical power of unused light. Note that the actual bandwidth of this filter is not crucial since the optical coherent detection is mode selective: only photons in the same spectral-temporal and polarization mode as the local oscillator will be detected. Nevertheless, the optical power of unused modes should be suppressed so that they are well below the power of the local oscillator.

\begin{figure}[t]
	\includegraphics[width=.45\textwidth]{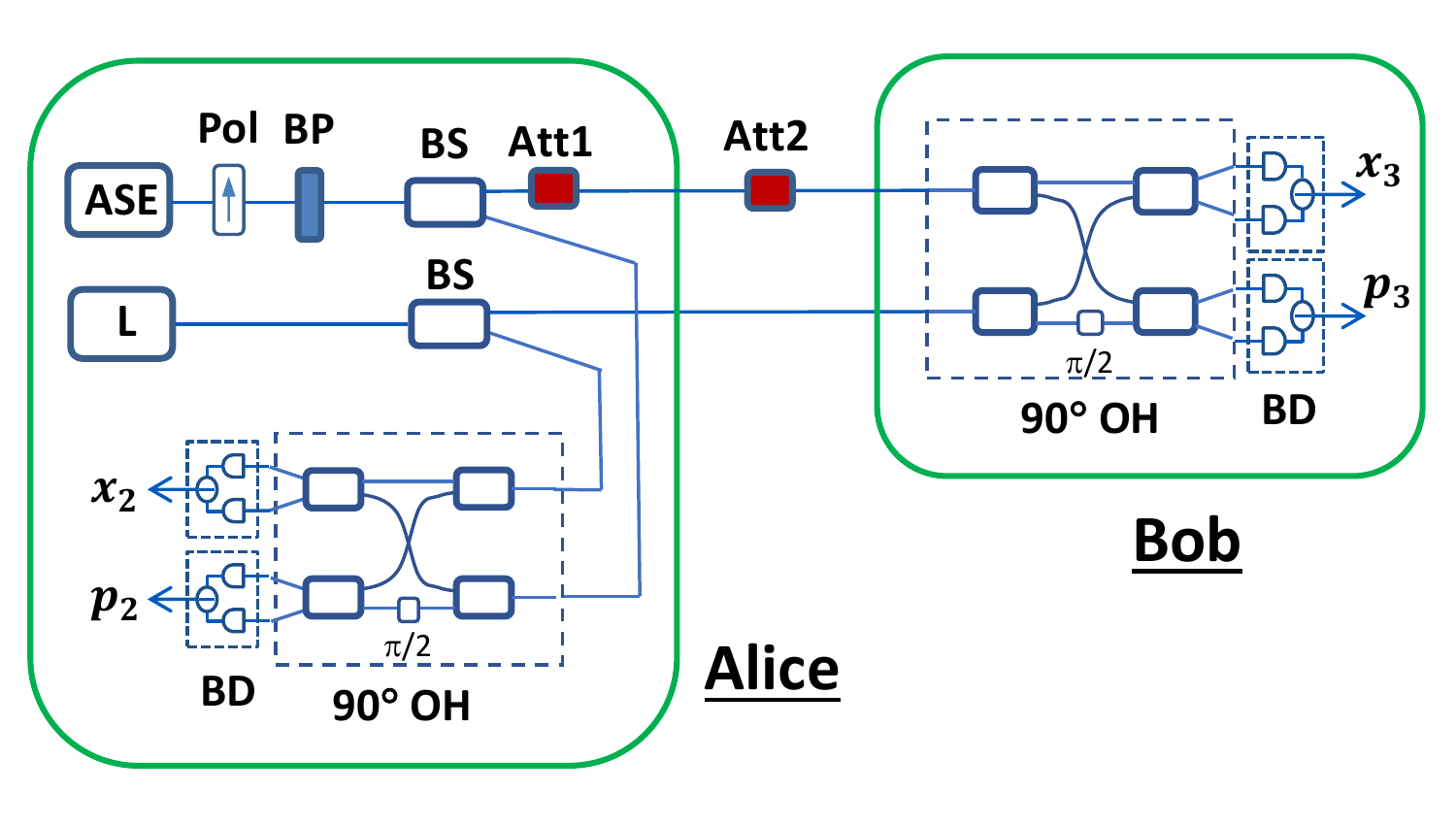}
	\captionsetup{justification=raggedright,
					singlelinecheck=false }
	\caption{Experimental setup. ASE, broadband amplified spontaneous emission source; L, narrow-band laser source as local oscillator; BP, optical band pass filter; BS, balanced beam splitter; Pol, fiber polarizer; Att1/Att2, variable optical attenuator; $90^\textup{o}$ OH, 90-degree optical hybrid; BD, balanced photodetector.} 
	\label{fig:2}
\end{figure}

The filtered thermal light is split by a balanced beam splitter into two modes: one is sent to Bob after passing through an optical attenuator while the other is measured locally by Alice using conjugate homodyne detection. Note that there are two optical attenuators shown in Fig. 2: Att1 represents a ``trusted'' attenuator inside Alice's system which cannot be controlled by Eve. It provides the desired attenuation $\eta_0$ shown in Eq. (1); Att2 represents the channel loss which is fully controlled by Eve. In this proof-of-principle experiment, a single optical attenuator is employed to provide the combined attenuation of Att1 and Att2. 

A cw laser source with a central wavelength of 1542nm (Clarity-NLL-1542-HP from Wavelength Reference) is employed to provide local oscillators for both Alice's and Bob's conjugate homodyne detection systems. For long distance applications, instead of transmitting a local oscillator from Alice to Bob, it is more appropriate to generate Bob's local oscillator locally \cite{Qi15, Soh15, Huang15}. The conjugate homodyne detection is implemented using a 90-degree optical hybrid and two balanced photodetectors. Limited by the detectors available, different types of balanced photodetectors are employed in Alice's and Bob's systems. The corresponding bandwidths are 100MHz (Alice) and 75MHz (Bob). The outputs of all the balanced photodetectors are sampled by a real time oscilloscope.

One important deviation of the setup shown in Fig. 2 from the ideal passive QKD protocol discussed in the previous Section is that a multi-mode (rather than a single mode) thermal source is employed in the actual experiment. On one hand, this modification greatly simplifies the implementation since it is experimentally challenging to prepare a single mode thermal state; on the other hand, the existence of unused modes could contribute additional noise, as we will discuss below.

Within the integration time of the homodyne detector, the output of the light source (even after the 0.8nm spectral filter) contains many spectral-temporal modes of independent thermal states. To generate correlated keys, Alice and Bob have to measure the same mode. In practice, the modes measured by Alice and Bob (which are determined by the modes of their local oscillators) may not be perfectly overlapped. We define the mode measured by Bob as mode $\vert B\rangle$. The mode measured by Alice may be decomposed as
\bes\label{eq2} \vert A\rangle=a \vert B\rangle + b \vert B'\rangle \ees
where $\vert B'\rangle$ represents the mode orthogonal to mode $\vert B\rangle$, and $\vert a \vert^2 + \vert b \vert^2 =1$. Without loss of generality, we assume the mode overlap coefficient $a$ is a real number.

\begin{figure}[t]
	\includegraphics[width=.5\textwidth]{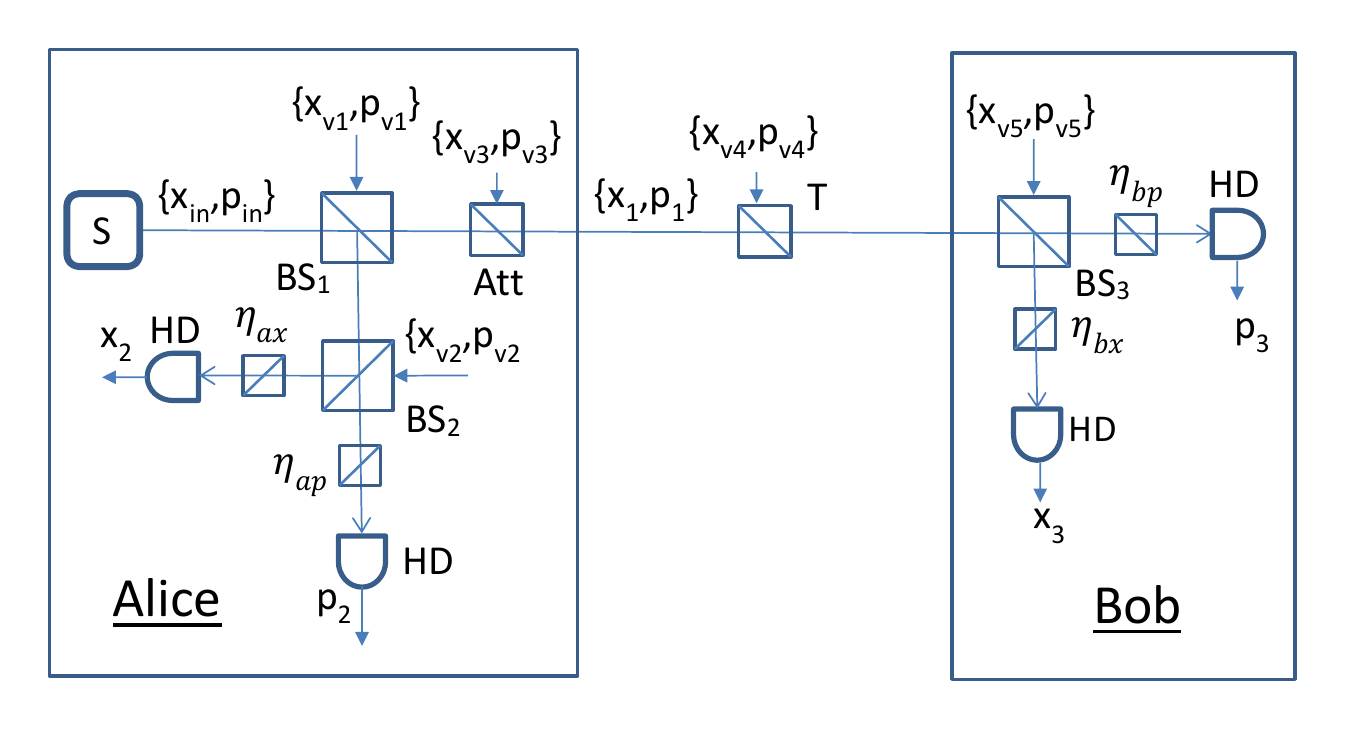}
	\captionsetup{justification=raggedright,
					singlelinecheck=false }
	\caption{Noise model of the experimental setup. S, thermal source; $\textup{BS}_{1-3}$, balanced beam splitter; Att, optical attenuator; HD, homodyne detector; T, a beam splitter emulating channel loss; $\eta_{ax}$, $\eta_{bx}$, $\eta_{ap}$, $\eta_{bp}$, beam splitters emulating losses of homodyne detectors.}
	\label{fig:3}
\end{figure}

For simplicity, we only consider the X-quadrature below. The P-quadrature can be studied in a similar way. The X-quadrature of the outgoing mode (see Fig. 3) is given by
\bes\label{eqa3} x_1=\sqrt{\frac{\eta_0}{2}}x_{in}-\sqrt{\frac{\eta_0}{2}}x_{v1}-\sqrt{1-\eta_0}x_{v3}, \ees 
where $x_{in}$ is the X-quadrature of mode $\vert B\rangle$ of the source, $\eta_0$ is the transmittance of the optical attenuator, $x_{v1}$ and $x_{v3}$ represent vacuum noises introduced by beam splitter one and the optical attenuator, respectively. 

Similarly, Alice's measurement result of the X-quadrature is given by
\begin{equation}
\begin{split}
&x_2=\dfrac{\sqrt{\eta_{ax}}}{2}(ax_{in}+bx'_{in})+\dfrac{\sqrt{\eta_{ax}}}{2}(ax_{v1}+bx'_{v1})\\
&+\sqrt{\frac{\eta_{ax}}{2}}x_{v2}-\sqrt{1-\eta_{ax}}x_{va}+E_{ax},
\end{split}
\end{equation}
where $\eta_{ax}$ and $E_{ax}$ are the efficiency and noise of Alice's homodyne detector for  X-quadrature measurement; $x'_{in}$ is the X-quadrature of mode $\vert B'\rangle$ of the source; $x'_{v1}$ is the vacuum noise in mode $\vert B'\rangle$ from beam splitter one; $x_{v2}$ and  $x_{va}$ are vacuum noises associated with beam splitter two and $\eta_{ax}$.

Given $x_2$, Alice's optimal estimation of $x_1$ is
\bes\label{eqa5} x_{opt}=\alpha_A x_2, \ees 
where $\alpha_A=\langle x_1 x_2 \rangle/\langle x_2^2 \rangle$ \cite{Poizat94, Grangier98}.

Using Eqs. (3) and (4), we can determine $\alpha_A$ as
\bes\label{eqa6} \alpha_A=\dfrac{n_0 a \sqrt{2\eta_0 \eta_{ax}}}{n_0\eta_{ax}+2\upsilon_{ax}+2}, \ees 
where $n_0$ is the average photon number per mode from the source and $\upsilon_{ax}=\langle E_{ax}^2 \rangle$ is the detector noise variance. The relation $\langle x_{in}^2 \rangle=\langle (x'_{in})^2 \rangle=2n_0+1$, which is a good approximation for the broadband light source used in our experiment, has been used in the above derivation.

Using Eqs. (3) to (6), Alice's uncertainty on $x_1$ given her measurement result of $x_2$ can be determined as
\begin{equation}
\begin{split}
&V_{x1|x2}=\langle (x_1-\alpha_A x_2)^2 \rangle=\\
&\dfrac{2V_A \eta_0 (\upsilon_{ax}+1)+V_A^2 \eta_{ax}(1-a^2)}{V_A \eta_{ax}+2\eta_0(\upsilon_{ax}+1)}+1,
\end{split}
\end{equation}
where $V_A=\eta_0 n_0$ is the equivalent modulation variance of the outgoing mode.

The excess noise due to the passive state preparation scheme is $\varepsilon_A=V_{x1|x2}-1$, which can be determined from Eq. (7) as
\bes\label{eq8} \varepsilon_A=\dfrac{2V_A \eta_0 (\upsilon_{ax}+1)+V_A^2 \eta_{ax}(1-a^2)}{V_A \eta_{ax}+2\eta_0(\upsilon_{ax}+1)}. \ees

Once the QKD excess noise has been appropriately quantified, we can apply the standard security proof for GMCS QKD to calculate the secure key rate. In next section, we will experimentally determine the mode mismatch $a$ and other system parameters to evaluate the performance of our system.

\section{Experimental results}
\label{sec:4}

In the experiment, both the broadband light source and the local oscillator are operated in cw mode. A 2GHz bandwidth real time oscilloscope is employed to sample the outputs of Alice and Bob's detectors. The modes sampled by Alice and Bob are determined by lengths of optical paths and also the electrical frequency responses of the optical homodyne detectors. Due to experimental imperfections, the modes measured by Alice and Bob are not perfectly overlapped, i.e., $a<1$. To determine $a$, we calculate the correlation between Alice and Bob's measurement results at different output photon numbers.

Alice's X-quadrature measurement result, $x_2$, is given by Eq. (4). Similarly, when no optical attenuation is applied ($\eta_0=1$, $T=1$, see Fig. 3), Bob's X-quadrature measurement result, $x_3$, is given by
\begin{equation}
\begin{split}
&x_3=\dfrac{\sqrt{\eta_{bx}}}{2}x_{in}-\dfrac{\sqrt{\eta_{bx}}}{2}x_{v1}\\
&+\sqrt{\frac{\eta_{bx}}{2}}x_{v5}-\sqrt{1-\eta_{bx}}x_{vb}+E_{bx},
\end{split}
\end{equation}
where $\eta_{bx}$ and $E_{bx}$ are the efficiency and noise of Bob's homodyne detector for  X-quadrature measurement; $x_{v5}$ and  $x_{vb}$ are vacuum noises associated with beam splitter 3 and $\eta_{bx}$ (see Fig. 3).

Using Eqs. (4) and (9), it is easy to show
\bes\label{eq10} \langle x_2^2 \rangle=\frac{\eta_{ax}}{2}n_0+\upsilon_{ax}+1, \ees 
\bes\label{eq11} \langle x_3^2 \rangle=\frac{\eta_{bx}}{2}n_0+\upsilon_{bx}+1, \ees 
\bes\label{eq12} \langle x_2 x_3 \rangle=\frac{\sqrt{\eta_{ax}\eta_{bx}}}{2}n_0a. \ees 

From Eqs. (10)-(12), the correlation coefficient between $x_2$ and $x_3$ is given by

\begin{equation}
\begin{split}
&Corr=\frac{\langle x_2 x_3 \rangle}{\sqrt{\langle x_2^2 \rangle \langle x_3^2 \rangle }}=\\
&\frac{n_0\sqrt{\eta_{ax}\eta_{bx}}}{\sqrt{(\eta_{ax}n_0+2\upsilon_{ax}+2)(\eta_{bx}n_0+2\upsilon_{bx}+2)}}a.
\end{split}
\end{equation}

Eq. (13) shows that as $n_0$ approaches infinity, the correlation coefficient approaches the mode overlap coefficient $a$. Thus, we can determine $a$ experimentally by measuring the correlation between Alice and Bob's measurement results at high output photon number.

\begin{figure}[t]
	\includegraphics[width=.45\textwidth]{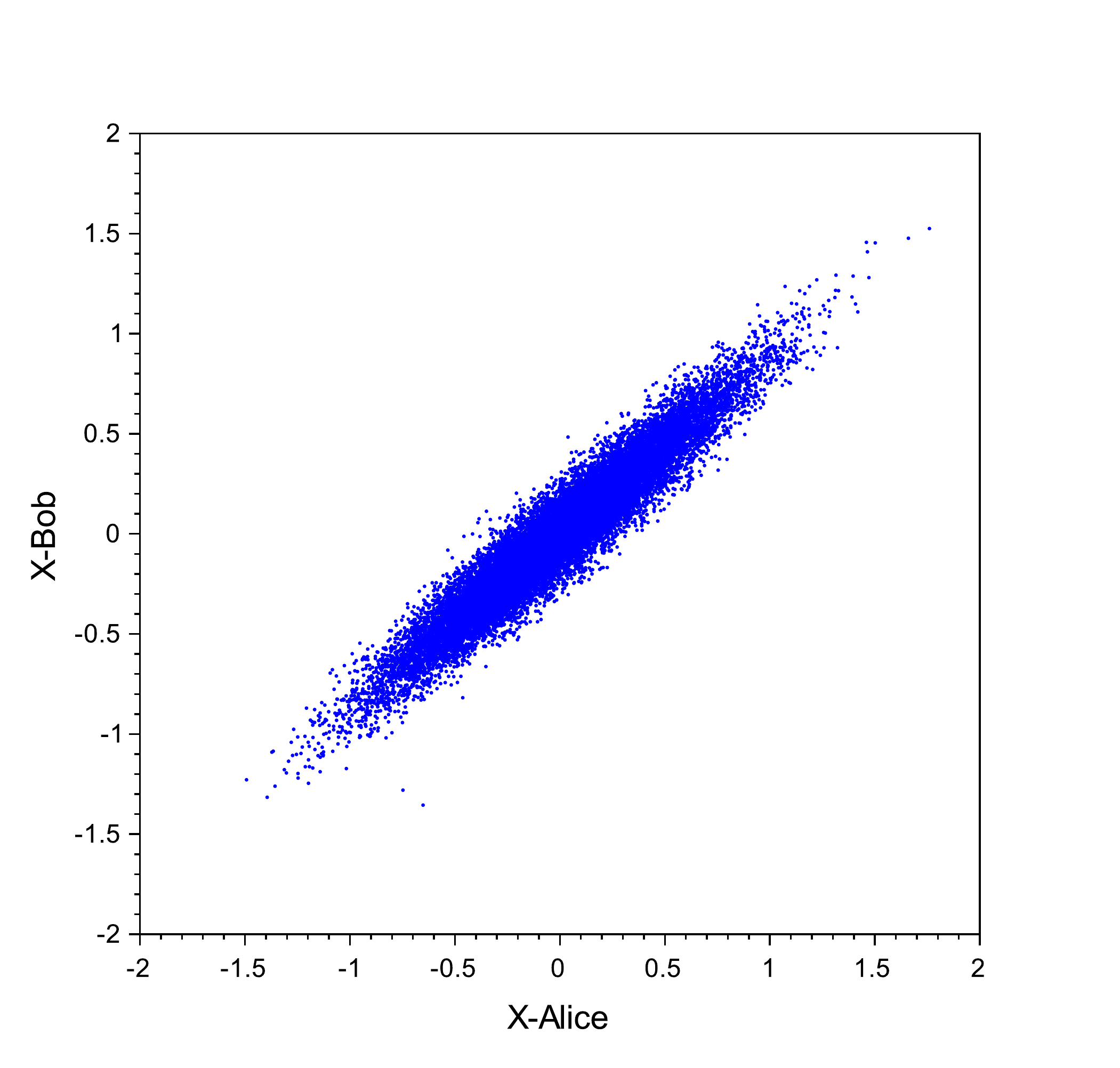}
	\captionsetup{justification=raggedright,
					singlelinecheck=false }
	\caption{Raw data of Alice and Bob's X-quadrature measurement results ($n_0$=880 and no optical attenuation applied).}
	\label{fig:4}
\end{figure}

\begin{figure}[t]
	\includegraphics[width=.45\textwidth]{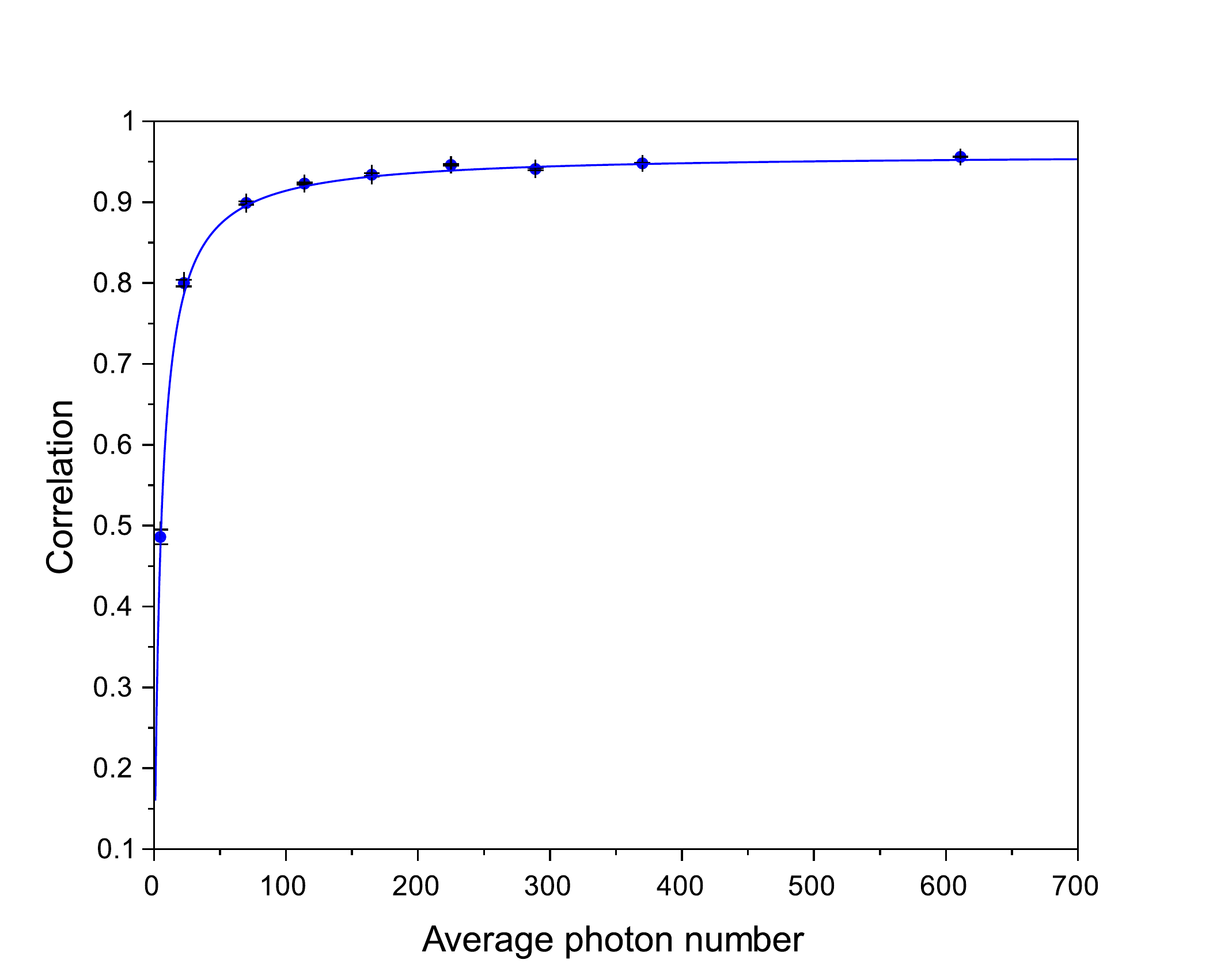}
	\captionsetup{justification=raggedright,
					singlelinecheck=false }
	\caption{The correlation coefficient of Alice and Bob's X-quadrature measurement results at different output photon numbers of the light source. Blue dots represent experimental results (error bars represent one standard deviation). The red line is a fitting curve using Eq. (13) with $a$ = 0.96.}
	\label{fig:5}
\end{figure}

We adjust $n_0$ by changing the pump power of the fiber amplifier and calculate the correlation coefficient from experimental data. In this experiment, no optical attention is applied ($\eta_{tot}=\eta_0 T=1$). The local oscillator power is 2mW. The detection efficiency and noise variance of Alice and Bob's conjugate homodyne detectors are $\eta_{ax}=0.43\pm0.01$, $\eta_{ap}=0.38\pm0.01$, $\eta_{bx}=0.54\pm0.01$, $\eta_{bp}=0.51\pm0.01$, $\upsilon_{ax}=0.17\pm0.01$, $\upsilon_{ap}=0.19\pm0.01$, $\upsilon_{bx}=0.24\pm0.01$,and $\upsilon_{ap}=0.23\pm0.01$. At each photon level, we collect 500,000 data samples, which are further divided into ten 50,000 data sets. The correlation coefficient is calculated for each data set.

Figure 4 shows the raw data of Alice and Bob's X-quadrature measurement results when the average photon number per mode from the source is $n_0=880$. Fig. 5 shows the correlation coefficient of Alice and Bob's X-quadrature measurement results at different $n_0$ (error bars represent one standard deviation). The red line is a curve fit using Eq. (13) with $a$ = 0.96 and detector parameters given above. The theory and experimental results match well.

To further justify the above noise model, we conduct experiments using a constant $n_0$ of 900 and different optical attenuation $\eta_{tot}$. This experiment can also be modeled using Eq. (13), with $\eta_{bx}$ replaced by $\eta_{tot} \eta_{bx}$. Note that when the optical loss is very high, the correlation between Alice and Bob's data is barely visible from the raw data, as evidenced by the case of $\eta_{tot}=-42.2$dB shown in Fig. 6. Nevertheless, by calculating the correlation coefficient from the raw data and comparing it with the theoretical prediction, the proposed noise model is well justified, as shown in Fig. 7.

\begin{figure}[t]
	\includegraphics[width=.45\textwidth]{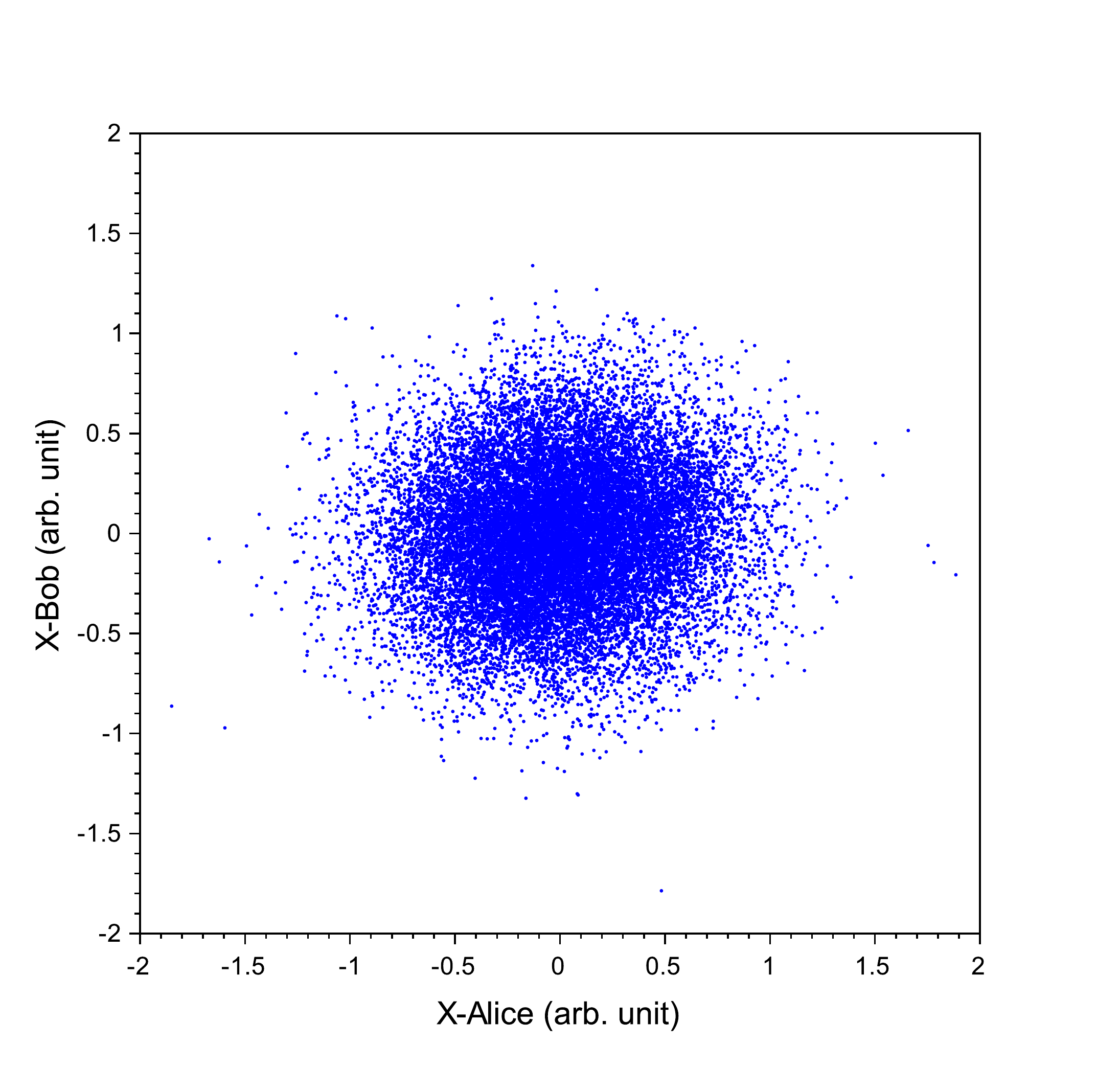}
	\captionsetup{justification=raggedright,
					singlelinecheck=false }
	\caption{Raw data of Alice and Bob's X-quadrature measurement results ($n_0=900$ and $\eta_{tot}=-42.2$dB).}
	\label{fig:6}
\end{figure}

\begin{figure}[t]
	\includegraphics[width=.45\textwidth]{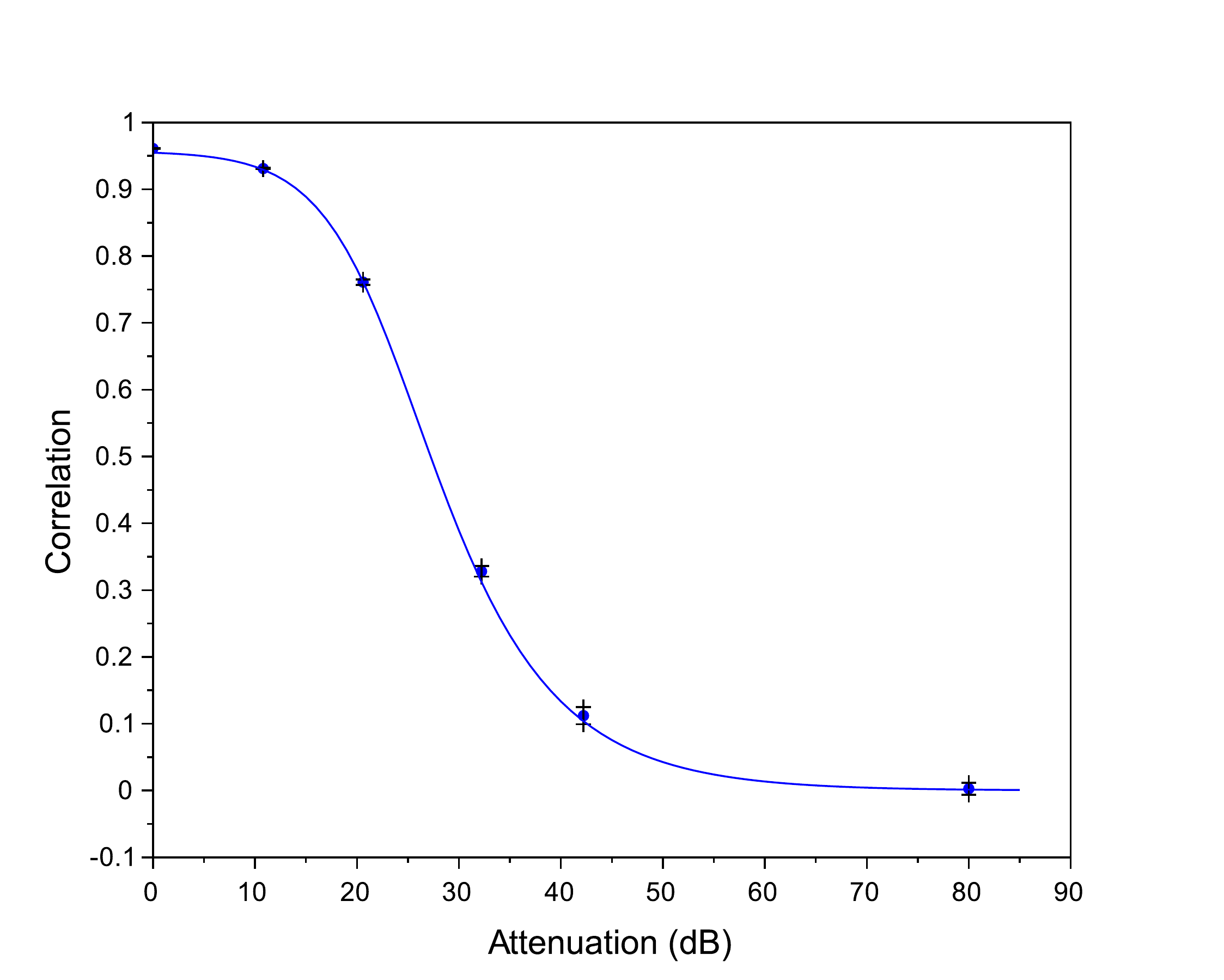}
	\captionsetup{justification=raggedright,
					singlelinecheck=false }
	\caption{The correlation coefficient of Alice and Bob's X-quadrature measurement results as a function of optical attenuation $\eta_{tot}=\eta_0 T$. Blue dots represent experimental results (error bars represent one standard deviation). The red line is a fit curve based on Eq. (13) by replacing $\eta_{bx}$ with by $\eta_{tot} \eta_{bx}$.}
	\label{fig:7}
\end{figure}

As we have shown in \cite{Qi18}, the well-established security proof of GMCS QKD can be applied to passive CV-QKD, as long as the excess noise of Alice is properly quantified using Eq. (8). We calculate the secure key as a function of the channel length using experimentally determined system parameters. Here, we assume the quantum channel is a single mode fiber with an attenuation coefficient of $\gamma=0.2$dB/km. As shown in Fig. 8, a practical distance above 80km could be achieved. The QKD distance could be further extended by improving the mode overlap factor $a$. Details of the secure key formulas are summarized in Appendix B.

To compare the simulated secure key rates with the experimental results shown in Fig. 7, we use Eq. (A11) in Appendix A to determine the mutual information between Alice and Bob from the experimentally determined correlation coefficient as $I_{AB}=\log_2\left( \dfrac{1}{1-Corr^2}\right)$. Two data points shown in Fig. 7, corresponding to $\eta_{tot}=-32.1$dB and $-42.2$dB, are used in the calculation. Since only one attenuator is applied in the experiment, we artificially divide the total attenuation $\eta_{tot}$ into Alice's attenuation $\eta_0$ and channel transmittance $T$ as $\lbrace \eta_0=0.0009, T=0.69\rbrace$ (for $\eta_{tot}=-32.1$dB) and $\lbrace \eta_0=0.0004, T=0.15\rbrace$ (for $\eta_{tot}=-42.2$dB). The calculated secure key rates are shown in Fig. 8 as two points, with error bars determined by one standard deviation of the correlation coefficient. 

\begin{figure}[t]
	\includegraphics[width=.45\textwidth]{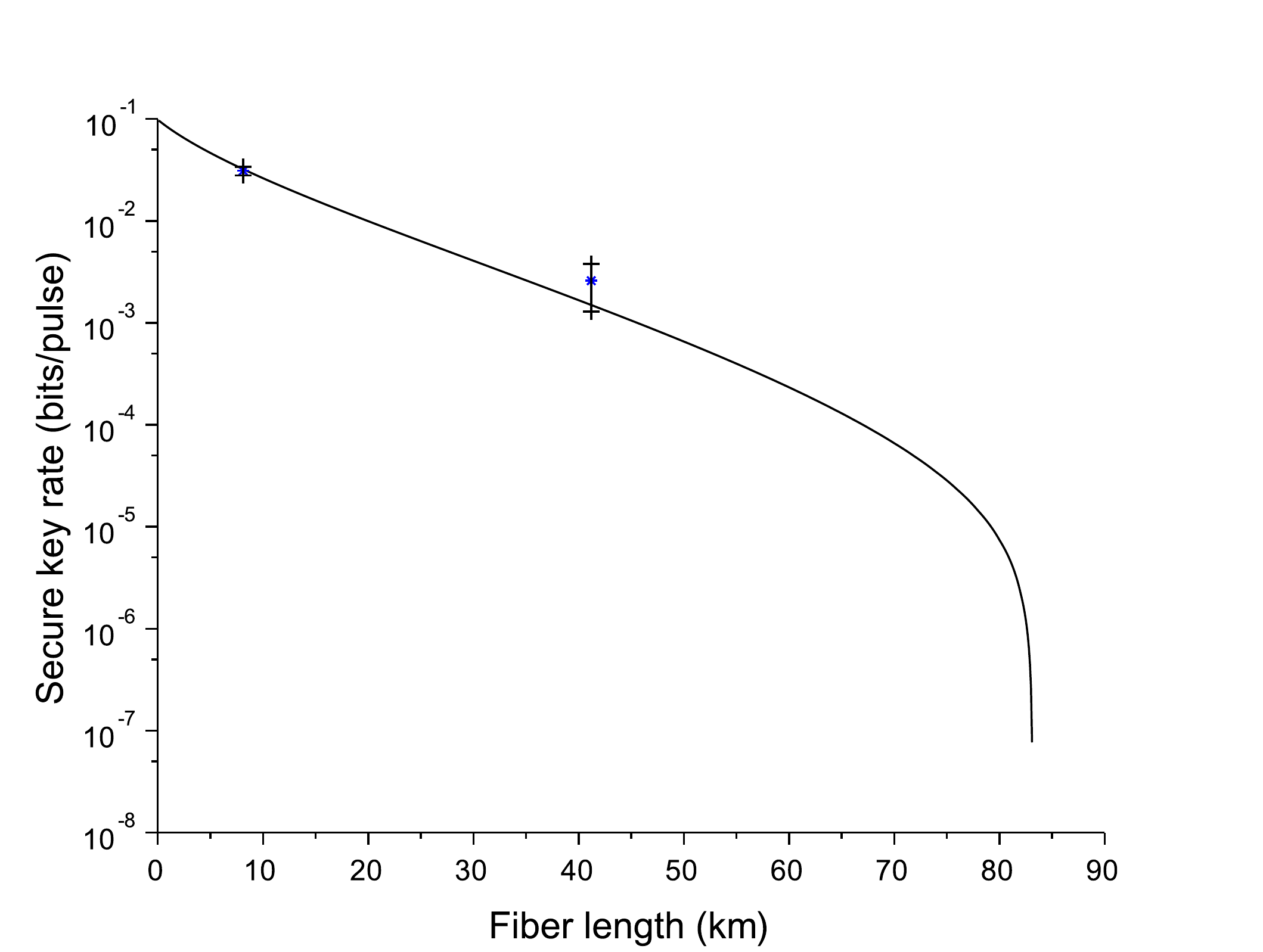}
	\captionsetup{justification=raggedright,
					singlelinecheck=false }
	\caption{The secure key rate as a function of channel loss. Simulation parameters: $a$ = 0.96, $n_0$=900, $\eta_{ax}=0.43$, $\eta_{ap}=0.38$, $\eta_{bx}=0.54$, $\eta_{bp}=0.51$, $\upsilon_{ax}=0.17$, $\upsilon_{ap}=0.19$, $\upsilon_{bx}=0.24$, $\upsilon_{ap}=0.23$. The efficiency of the reconciliation algorithm is $f=0.95$ (see details in Appendix B). The two data points are secure key rates calculated using the experimental results shown in Fig. 7 (error bars are determined by one standard deviation of the correlation coefficient).}
	\label{fig:8}
\end{figure}

\section{Summary}
\label{sec:5}

One common question on the passive CV-QKD scheme is whether we can trust the randomness from a thermal source. It is a common practice to apply a quantum random number generator \cite{Ma16, Herrero17} in prepare-and-measure QKD for state preparation and/or measurement basis selection. As we have discussed in \cite{Qi17}, while quantum randomness is ultimately connected to quantum superposition states, in the fully trusted device scenario, the quantum state received by the detector does not need to be a pure state. One illustrative example is the first quantum random number generator, where electrons from a radioactive source such as $^{90}\textup{Sr}$ are detected by a Geiger Mueller tube at random times \cite{Isida56, Schmidt70}. In this process, while the whole system (the radioactive nuclei and electrons) is in pure state, the state received by the detector is a mixture of 0-electron and 1-electron emission states. True randomness can be generated as long as Eve cannot access (or control) the radioactive source. Similar arguments can also be applied to randomness generated from spontaneous emission using a thermal source.

The security of QKD is only as good as its underlying assumptions \cite{Qi072}. In this paper, we have adopted a commonly used assumption in QKD that the QKD systems employed by Alice and Bob are fully trusted and cannot be accessed by Eve. In practice, any real-life QKD systems cannot be perfect. It is thus important to scrutinize all of the implementation details to identify potential side channels and develop the corresponding countermeasures. The investigation of loopholes and countermeasures in practical QKD systems plays a complementary role to security proofs.

In summary, we conduct experimental studies on the recently proposed passive CV-QKD protocol \cite{Qi18}, which is appealing for chip-scale implementation. When implemented with a practical multi-mode thermal source, one important issue is how to determine the excess noise contributed by photons in the unwanted modes. In this paper, we develop a noise model based on a practical setup, and conduct experiments to verify the above model using a commercial off-the-shelf amplified spontaneous emission source. Our results suggest that passive CV-QKD could be a cost-effective solution for metro-area QKD. 

This work was performed at Oak Ridge National Laboratory (ORNL), operated by UT-Battelle for the U.S. Department of Energy (DOE) under Contract No. DE-AC05-00OR22725. The authors acknowledge support from DOE Office of Cybersecurity Energy Security and Emergency Responce (CESER) through the Cybersecurity for Energy Delivery Systems (CEDS) program. 

\appendix

\section{Detailed noise model analysis}

In this appendix, we present a detailed noise model of the passive CV-QKD scheme taking into account imperfections of practical homodyne detectors. To gain some intuition on its security, we also study the well-known beam-splitting attack where Eve replaces the lossy quantum channel by a beam splitter with a suitable transmittance and measures both the X-quadrature and P-quadrature of the reflected light, as shown in Fig. 9. Note that in our model, we assume Eve's detectors are perfect while Alice and Bob's detectors are lossy and noisy.

\begin{figure}[t]
	\includegraphics[width=.5\textwidth]{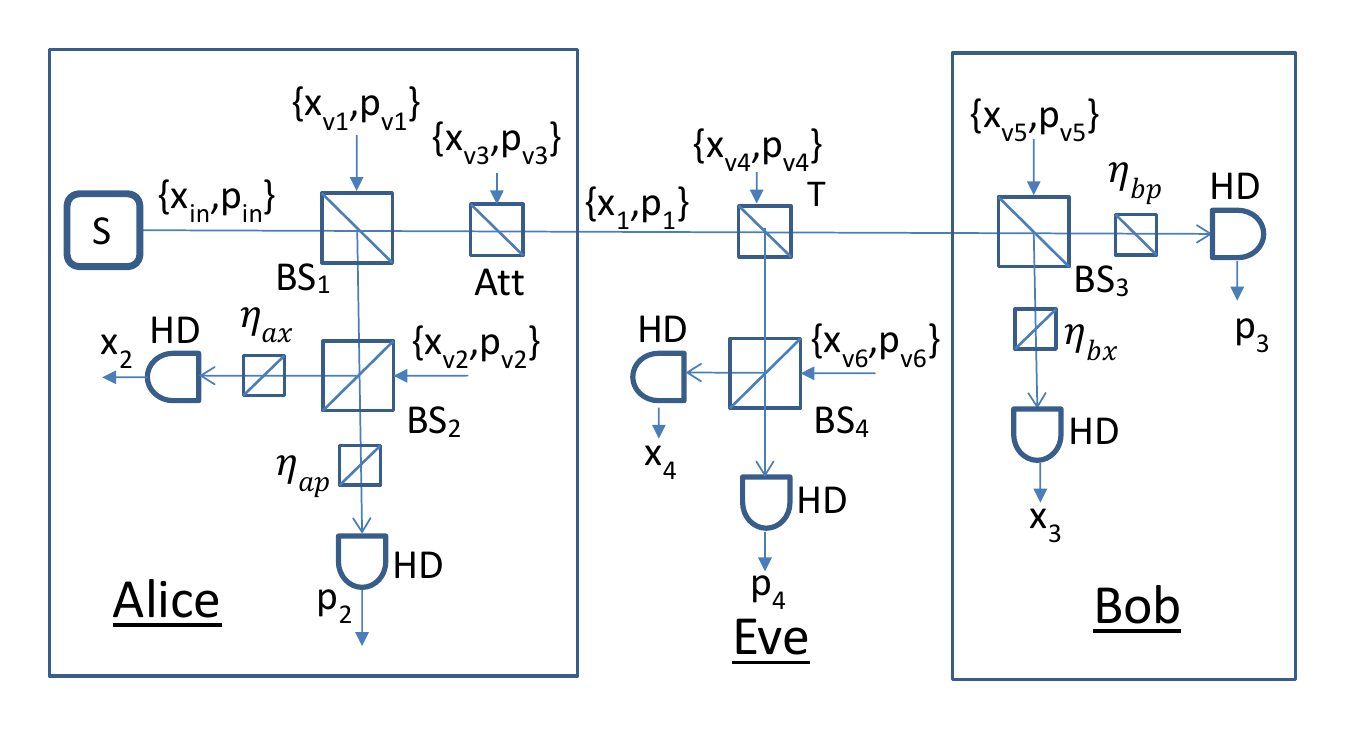}
	\captionsetup{justification=raggedright,
					singlelinecheck=false }
	\caption{Detailed noise model under the beam-splitting attack. S, thermal source; $\textup{BS}_{1-4}$, balanced beam splitter; Att, optical attenuator; HD, homodyne detector; T, a beam splitter emulating channel loss; $\eta_{ax}$, $\eta_{bx}$, $\eta_{ap}$, $\eta_{bp}$, beam splitters emulating losses of homodyne detectors.}
	\label{fig:9}
\end{figure}

For simplicity, we only consider the X-quadrature below. The P-quadrature can be studied in a similar way. The X-quadrature of the optical mode output from Alice (see Fig. 9) is given by
\bes\label{eqa1} x_1=\sqrt{\frac{\eta_0}{2}}x_{in}-\sqrt{\frac{\eta_0}{2}}x_{v1}-\sqrt{1-\eta_0}x_{v3}, \ees 
where $x_{in}$ is the X-quadrature of the output of the source, $\eta_0$ is the transmittance of the optical attenuator, $x_{v1}$ and $x_{v3}$ represent vacuum noises introduced by beam splitter one and the optical attenuator, respectively (see Fig. 9). Note the variance of vacuum noise is one in shot noise units, i.e., $\langle x_{v1}^2 \rangle=\langle x_{v3}^2 \rangle=1$.

Given the source outputs a thermal state with an average photon number of $n_0$, it is easy to show that the outgoing mode from Alice is also a thermal state with an average photon number of $\eta_0 n_0/2$. The output state from Alice in the passive CV-QKD is identical to that of GMCS QKD using active modulation with a modulation variance of
\bes\label{eqa2} V_A=\eta_0 n_0. \ees 

As we discussed in Sec. II, the secure key rate of CV-QKD is sensitive to excess noises. It is thus important to quantify how much excess noise is introduced by the passive state preparation scheme.

As shown in Fig. 9, Alice splits the output of the source into two modes using a balanced beam splitter, and measures the X-quadrature and P-quadrature of the local mode. Alice's X-quadrature measurement result is
\begin{equation}
\begin{split}
&x_2=\dfrac{\sqrt{\eta_{ax}}}{2}x_{in}+\dfrac{\sqrt{\eta_{ax}}}{2}x_{v1}\\
&+\sqrt{\frac{\eta_{ax}}{2}}x_{v2}-\sqrt{1-\eta_{ax}}x_{va}+E_{ax},
\end{split}
\end{equation}
where $\eta_{ax}$ and $E_{ax}$ are the efficiency and noise of Alice's homodyne detector for  X-quadrature measurement, $x_{v2}$ and $x_{va}$ are vacuum noises associated with beam splitter two and $\eta_{ax}$.

Given $x_2$, Alice's optimal estimation of $x_1$ is $x_{opt}=\alpha_0 x_2$, where $\alpha_0=\langle x_1 x_2 \rangle/\langle x_2^2 \rangle$ \cite{Poizat94, Grangier98}.

Using Eqs. (A1) and (A3), we can determine $\alpha_0$ as
\bes\label{eqa4} \alpha_0=\dfrac{n_0 \sqrt{2\eta_0 \eta_{ax}}}{n_0\eta_{ax}+2\upsilon_{ax}+2}. \ees 

In the above derivation, we define the detector noise variance as $\upsilon_{ax}=\langle E_{ax}^2 \rangle$. We also used the relation of $\langle x_{in}^2 \rangle=2n_0+1$, which is true for a thermal state.

Alice's uncertainty on $x_1$ given her measurement result of $x_2$ is $\Delta=V_{x1|x2}=\langle (x_1-\alpha_0 x_2)^2 \rangle$. Using Eqs. (A1) to (A4), we can determine the excess noise of state preparation as
\bes\label{eqa5} \varepsilon_A=\Delta-1=\dfrac{2V_A(\upsilon_{ax}+1)}{V_A \eta_{ax}+2\eta_0(\upsilon_{ax}+1)}\eta_0. \ees 

Equation (A5) suggests that the excess noise of the passive state preparation scheme can be effectively reduced by introducing large optical attenuation at Alice (i.e. $\eta_0 <<1$). Note to keep a desired variance $V_A$ of the outgoing mode, a large $n_0$ is required from the source to compensate a small $\eta_0$.

To further illustrate the role of the optical attenuation, we calculate the mutual information between Alice and Bob ($I_{AB}$), and the one between Eve and Bob ($I_{BE}$), under the beam-splitting attack. We emphasis that in this paper, we consider the CV-QKD protocol using heterodyne detection and reverse reconciliation, where Bob measures both the X-quadrature and P-quadrature of the incoming mode, and Alice tries to guess Bob's measurement results from her data. This is in contrast to the direct reconciliation protocol where Bob tries to determine Alice's data.

As shown in Fig. 9, a lossy channel with a transmittance of $T$ can be modeled by a beam splitter with an appropriate splitting ratio. In the beam splitting attack, the transmitted and reflected mode are detected by Bob and Eve respectively.

From Fig. 9, Bob's X-quadrature measurement result is given by
\begin{equation}
\begin{split}
&x_3=\dfrac{\sqrt{\eta_0 T \eta_{bx}}}{2}x_{in}-\dfrac{\sqrt{\eta_0 T \eta_{bx}}}{2}x_{v1}-\sqrt{\frac{(1-\eta_0) T \eta_{bx}}{2}}x_{v3}\\
&-\sqrt{\frac{(1-T) \eta_{bx}}{2}}x_{v4}+\sqrt{\frac{\eta_{bx}}{2}}x_{v5}-\sqrt{1-\eta_{bx}}x_{vb}+E_{bx},
\end{split}
\end{equation}
where $\eta_{bx}$ and $E_{bx}$ are the efficiency and noise of Bob's homodyne detector for X-quadrature measurement, $x_{vb}$ represents vacuum noise associated $\eta_{bx}$, vacuum noises introduced by other components are shown in Fig. 9.

Similarly, Eve's X-quadrature measurement result is given by
\begin{equation}
\begin{split}
&x_4=\dfrac{\sqrt{\eta_0 (1-T)}}{2}x_{in}-\dfrac{\sqrt{\eta_0 (1-T)}}{2}x_{v1}\\
&-\sqrt{\frac{(1-\eta_0) (1-T)}{2}}x_{v3}+\sqrt{\frac{T}{2}}x_{v4}+\dfrac{1}{\sqrt{2}}x_{v6},
\end{split}
\end{equation}

In reverse reconciliation, Alice tries to guess Bob's measurement result $x_3$ given her own measurement result $x_2$. Alice's optimal estimation of $x_3$ is $x_A=\alpha_1 x_2$, where $\alpha_1=\langle x_3 x_2 \rangle/\langle x_2^2 \rangle$. From Eqs. (A3) and (A6), Alice's conditional variance on Bob's measurement result $V_{B|A}$ can be determined as:
\begin{equation}
\begin{split}
&V_{B|A}=\langle (x_3-\alpha_1 x_2)^2 \rangle=\langle x_3^2 \rangle-\dfrac{\langle x_2 x_3 \rangle ^2}{\langle x_2^2 \rangle}\\
&=\dfrac{V_A T \eta_{bx}(\upsilon_{ax}+1)}{\frac{\eta_{ax}}{\eta_0}V_A+2\upsilon_{ax}+2}+\upsilon_{bx}+1.
\end{split}
\end{equation}

Since secure key could be generated from both X and P, the mutual information between Alice and Bob is given by
\bes\label{eqa9} I_{AB}=\log_2\left( \dfrac{V_B}{V_{B|A}}\right), \ees 
where Bob's X-quadrature variance $V_B$ is given by
\bes\label{eqa10} V_B= \langle x_3^2 \rangle=\dfrac{V_A T \eta_{bx}}{2}+\upsilon_{bx}+1. \ees 

Note that by using Eqs. (A8) and (A10), we can write Eq. (A9) as
\bes\label{eqa11} I_{AB}=\log_2\left( \dfrac{1}{1-Corr^2}\right), \ees 
where $Corr=\frac{\langle x_2 x_3 \rangle}{\sqrt{\langle x_2^2 \rangle \langle x_3^2 \rangle }}$ is the correlation coefficient. Eq. (A11) allows us to use the experimentally determined correlation coefficient to estimate the mutual information between Alice and Bob.

Following a similar procedure, Eve's conditional variance on Bob's measurement result and the mutual information between Eve and Bob can be determined as:
\bes\label{eqa12} V_{B|E}=\dfrac{V_A T \eta_{bx}}{V_A(1-T)+2}+\upsilon_{bx}+1, \ees
and 
\bes\label{eqa13} I_{BE}=\log_2\left(\dfrac{V_B}{V_{B|E}}\right). \ees 

From Eqs. (A8) to (A13), for a given $V_A$, we can effectively suppress $V_{B|A}$ and thus increase $I_{AB}$ by reducing the value of $\eta_0$. In contrast, $I_{BE}$ is independent of the value of $\eta_0$. This shows the information advantage between Alice and Bob over that between Eve and Bob can be improved by introducing optical loss at Alice. 

In principle, secure key could be generated as long as $V_{B|A}<V_{B|E}$. The corresponding constraint on $\eta_0$ can be determined from Eqs. (A8) and (A12) as
\bes\label{eqa14} \eta_0 < \dfrac{\eta_{ax}}{(\upsilon_{ax}+1)(1-T)}. \ees

\section{Secure key rate calculation}

The asymptotic secure key rate of GMCS QKD, in the case of reverse reconciliation, is given by Refs.~\cite{Lodewyck07, Fossier09} 
\bes\label{b1} R=fI_{AB}-\chi_{BE}, \ees
where $I_{AB}$ is the Shannon mutual information between Alice and Bob; $f$ is the efficiency of the reconciliation algorithm; $\chi_{BE}$ is the Holevo bound on information between Eve and Bob. 

For an optical fiber link with an attenuation coefficient of $\gamma$, the channel transmittance is given by
\bes\label{b2} T=10^{\frac{-\gamma L}{10}},\ees
where $L$ is the fiber length in kilometers.

In the case of conjugate homodyne detection, the X-quadrature noise added by Bob's detector (referred to Bob's input) is given by \cite{Fossier09}
\bes\label{b3} \chi_{het}=[1+(1-\eta_{bx})+2\upsilon_{bx}]/\eta_{bx}.\ees

The channel-added noise (referred to the channel input) is given by
\bes\label{b4} \chi_{line}=\frac{1}{T}-1+\varepsilon_A,\ees
where $\varepsilon_A$ is defined in Eq. (8) in the main text.

The overall noise is given by
\bes\label{b5} \chi_{tot}=\chi_{line}+\dfrac{\chi_{het}}{T}. \ees

Since both quadratures can be used to generate secure key, the mutual information between Alice and Bob can be determined by
\bes\label{b6} I_{AB}=\log_2\dfrac{V+\chi_{tot}}{1+\chi_{tot}}, \ees
where $V=V_A+1$.

To estimate $\chi_{BE}$, we assume that Eve cannot control the imperfections in Bob's system. This noise model has been widely used in CV-QKD experiments \cite{GMCSQKD, Jouguet13, Kumar15, Qi07, Lodewyck07, Huang16, ZLC19}. Under this model, the Holevo bound of the information between Eve and Bob is given by \cite{Lodewyck07} 
\bes\label{b7} \chi_{BE}=\sum_{i=1}^2 G\left( \dfrac{\lambda_i-1}{2} \right) - \sum_{i=3}^5 G\left( \dfrac{\lambda_i-1}{2}\right), \ees
where $G(x)=(x+1){\rm{log}}_2(x+1)-x{\rm{log}}_2x$.

\bes\label{b8} \lambda_{1,2}^2=\frac{1}{2} \left[ A\pm \sqrt{A^2-4B} \right], \ees
where
\bes\label{b9} A=V^2 (1-2T)+2T+T^2 (V+\chi_{line})^2, \ees
\bes\label{b10}B=T^2(V\chi_{line}+1)^2. \ees
 
\bes\label{b11} \lambda_{3,4}^2=\frac{1}{2} \left[ C\pm \sqrt{C^2-4D} \right], \ees
where
\begin{equation}
\begin{split}
C=\dfrac{1}{(T(V+\chi_{tot}))^2} [ A\chi_{het}^2+B+1+2\chi_{het} \\
( V\sqrt{B}+T(V+\chi_{line})) +2T(V^2-1)],
\end{split}
\end{equation}

\bes\label{b13}D=\left( \dfrac{V+\sqrt{B}\chi_{het}}{T(V+\chi_{tot})} \right) ^2. \ees 
\bes\label{b14} \lambda_5=1. \ees

We remark that in this paper, all the excess noise generated at Alice are assumed to be caused by Eve's attack. This is a very conservative assumption. An alternative way to deal with the excess noise is to develop a suitable ``trusted'' noise model \cite{Usenko16}, which may lead to a better secure key rate.

\end{document}